\def\Version{ 2.07
    }

  %%%%%%%%%%%%%%%%%%%%%%%%%%%%%%%%%%%%%%%%%%%%%%%%%%%
  %%%  PROCESS THIS FILE WITH `tex' and `dvips'  %%%%
  %%%%%%%%%%%%%%%%%%%%%%%%%%%%%%%%%%%%%%%%%%%%%%%%%%%

 %%%%%%%%%%%%%%%%%%%%%%%%%%%%%%%%%%%%%%%%%%%%%%%%%%%%%%%%%%%%%%%%%%%%%%%%%%%%%%%%%
 % IF DESIRED ALSO COMMENT OUT \PrintVersionNumber, \PrintTimestamp farther BELOW 
 %%%%%%%%%%%%%%%%%%%%%%%%%%%%%%%%%%%%%%%%%%%%%%%%%%%%%%%%%%%%%%%%%%%%%%%%%%%%%%%%% 

% Outlining is at end!

%: Page Setup for 8.5 x 11 inch paper        

\message{ Assuming 8.5" x 11" paper }    

\magnification=\magstep1	          % \magstep1=1200
%% \mag \magstep1                            %

\raggedbottom

\parskip=9pt

% \hsize=6.4 true in
% \vsize=8.7 true in
%
% \hoffset=0.27 true in
% \voffset=0.28 true in

\def\singlespace{\baselineskip=12pt}      % spacing for stuff like abstract
\def\sesquispace{\baselineskip=16pt}      % spacing for main text
\def\sesquispace{\baselineskip=20pt}      % draft spacing %

%: Load or define some TeX macros            
%:  general purpose macro files              

%% \input eplain   % [not installed on mars apparently]

%% Incorporate these files here, when paper finished
%\input mathmacros
\def\ideq{\equiv}		% triple equal sign
\def\eps{\varepsilon}           % ok??

% \input mathmacros.greekbold
% \input msmacros
% %~%~%~%~%~%~%~%~%~%~%~%  START of the file `msmacros' %~%~%~%~%~%~%~%~%~%~%~%~%~%~%~%~%~  

%   (Outline* "\\" 1  "%%" 1)

% This is `msmacros.tex' TimeStamp{2017-Feb-27 01:54:18 22707.52506 at elvis}
%
% The macros here are specifically for manuscripts, and concern such things as
% title, abstract, section titles, page formatting, etc.

% For Spanish accents and umlauts do `tengeneza-tex-acentos'

% Provisionally handling subsection titles as in hamilton-crew paper.  
% Have also taken title-font and sec headings from there. 

%%
\let\miguu=\footnote
\def\footnote#1#2{{$\,$\parindent=9pt\baselineskip=13pt%
\miguu{#1}{#2\vskip -7truept}}}

\def\linebreak{\hfil\break}

\def\pagebreak{\vfil\break}

	% to prevent page break (usually works)

\def\BulletItem #1 {\item{$\bullet$}{#1 }}
\def\bulletitem #1 {\BulletItem{#1}}

%% Following is for extended quotations, it single-spaces and narrows the text. 
% Amalgamate it with \BeginIndentation ?
% It seems to work tolerably well (see valdivia paper and tetralemma paper)

 % The initial smallskip is needed to activate the rest, apparently!

 % \def\LEMMA{\smallskip\noindent {\csmc Lemma }}
 % \def\LEMMA{\noindent {\csmc Lemma \ }}

\def\PrintVersionNumber{
 \vskip -1 true in \medskip 
 \rightline{version \Version} 
 \vskip 0.3 true in \bigskip \bigskip}

\def\author#1 {\medskip\centerline{\it #1}\bigskip}

\def\address#1{\centerline{\it #1}\smallskip}

\def\furtheraddress#1{\centerline{\it and}\smallskip\centerline{\it #1}\smallskip}

\def\email#1{\smallskip\centerline{\it address for email: #1}} 

\def\AbstractBegins
{
 \singlespace                                        % spacing for abstract
 \bigskip\leftskip=1.5truecm\rightskip=1.5truecm     % begin indentation
 \centerline{\bf Abstract}
 \smallskip
 \noindent	% this doesn't seem to take effect over a blank line
 } 
\def\AbstractEnds
{
 \bigskip\leftskip=0truecm\rightskip=0truecm       %  end indentation
 }

\def\section #1 {\bigskip\noindent{\headingfont #1 }\par\nobreak\smallskip\noindent}

\def\subsection #1 {\medskip\noindent{\subheadfont #1 }\par\nobreak\smallskip\noindent}
 %
 % see alternative defs in: ms/texfiles/developing.macros/warehoused.macros

\def\ReferencesBegin
{
 \singlespace					   % single spacing
 \vskip 0.5truein
 \centerline           {\bf References}
 \par\nobreak
 \medskip
 \noindent
 \parindent=2pt
 \parskip=6pt			% earlier was 10 pt and then 4pt
 }
 %
 % Don't use \raggedright here !
 % It got even more overfull hboxes, and it ruined the
 % right-justification of the references!

\def\reference{\hangindent=1pc\hangafter=1} % to put before each reference
% try 2pc here (what's a pc anyhow? perhaps a pico?) %%%

\def\ref{\reference}

 %
 % To separate individual items within a single reference

\def\journaldata#1#2#3#4{{\it #1\/}\phantom{--}{\bf #2$\,$:} $\!$#3 (#4)}
 %
 % Arguments are {journal name}{volume}{pages}{year}
 % (volume can also be {volume plus issue number})
 % A space after #1 seems to do something

 %
 % Probably works with both old- and new-style arxiv identifiers, if use as follows:
 %   papers having old identifier: \arxiv{gr-qc/9909123}
 %   papers having new identifier: \arxiv{0709.1234}

%% Macros for my own homepage 
% First the PI versions, which are free of trouble, then lo demas

\def\webhome{{\tt http://www.pitp.ca/personal/rsorkin/}}
 % \def\webhome{{\tt http://www.perimeterinstitute.ca/personal/rsorkin/}}
 %
 % My own homepage 

 % My own webpage with my papers
 %
 % http://www.perimeterinstitute.ca/personal/rsorkin/some.papers/

%% SU versions and related -- largely superfluous now!
% In the following, all the messiness is caused by the tilde in my SU web address.  

 % \def\webhome{{\tt {http://www.physics.syr.edu/}{\webtilde}{sorkin/}}}

\def\webtilde{\lower2pt\hbox{${\widetilde{\phantom{m}}}$}}
 % Only needed for my web page at SU

 % \def\webhome{{\tt {http://www.physics.syr.edu/}{\webEscapedTilde}{sorkin/}}}
 %
 % Oficially preferred form without tilde (but with percent sign!)

\def\hpfll#1{\webhome{\tt{lisp.library/}}}
 %
 % another, similar kluge!

%% Some provisional fonts
% In these fonts, maybe better to scale 2 1 1 rather than 2 1 0 (manake?)
% magstep1 means 12 pt, magstep1 2 means 14.4 pt 

\font\titlefont=cmb10 scaled\magstep2 

\font\headingfont=cmb10 at 12pt
 %
 % \font\headingfont=cmb10 at 13pt
 % \font\headingfont=cmb10 scaled\magstep1
 %% magstep2 is too big. Is magstep1 big enough? (if not, try 13pt)

\font\subheadfont=cmssi10 scaled\magstep1 % sans serif italic
 %
 %\font\subheadfont=cmb10 scaled\magstep0

  % caps and small caps 

% Next commented out, because it's trouble when the file "msmacros" is
% included bodily in some other tex file! 
% \endinput   

% %~%~%~%~%~%~%~%~%~%~%~%  END of the file `msmacros' %~%~%~%~%~%~%~%~%~%~%~%~%~%~%~%~%~  

% \input figure.macros  % basically miniltx

%:  possibly load { miniltx } and/or { graphicx } here 

% \input graphicx                 % this inputs miniltx
% \input miniltx

%:  macro for figure caption (EXPERIMENTAL)  

%:  ad hoc macros for this paper             

\def\ave#1{\langle \, #1 \, \rangle}
\def\betabar{\bar\beta}
\def\Amoja{{A1}}
\def\A2{{A2}}

%: phantom input	   	       	     

% Not clear why, but following somehow helps with layout of first page:

\phantom{}

%: Should we print a version number ior timestamp?

\PrintVersionNumber   % [[ comment out? ]]
%% \PrintTimestamp       % comment out?

%: Title                                     

%% See ~/ms/hamilton.crew/working.eltex for example of footnote-style addresses

\sesquispace
\centerline{{\titlefont An equation for a time-dependent profit rate}\footnote{$^{^{\displaystyle\star}}$} 
%% \centerline{{\titlefont An Upper Bound on the Profit Rate}
%% \centerline{{\titlefont FINAL LINE OF TITLE}\footnote{$^\star$}%
%
{The present article is a revised version of a 1982 manuscript that was submitted
 unsuccessfully to various economics journals.  The main equations and results are the same
 as before, with some improvements and changes of notation.  The surrounding discussion has
 been revised significantly, especially the concluding section.
%% [[To appear in ???.]]  
%% \eprint{arxiv yymmnnn}
}}
%%%%%%%%%%%%%%%%%%%%%%%%%%%%%%%%%%%%%%%%%%%%%%%%%%%%%%%%%%%%%%%%%%%%%%%%%%%
%%%%%%%      Enclosing \footnote ``within'' \centerline     %%%%%%%%%%%%%%%
%%%%%%%      is essential to get the asterisk right!        %%%%%%%%%%%%%%% 
%%%%%%%%%%%%%%%%%%%%%%%%%%%%%%%%%%%%%%%%%%%%%%%%%%%%%%%%%%%%%%%%%%%%%%%%%%%
%%%%%%%	     ADVERTENCIA  Even if a line of the title runs  %%%%%%%%%%%%%%% 
%%%%%%%      over you DON'T get an error message (why??)    %%%%%%%%%%%%%%% 
%%%%%%%%%%%%%%%%%%%%%%%%%%%%%%%%%%%%%%%%%%%%%%%%%%%%%%%%%%%%%%%%%%%%%%%%%%%

\bigskip

%: Authors                                   

\singlespace			        % spacing for addresses etc.

\author{Rafael D. Sorkin}
\address
 {Perimeter Institute, 31 Caroline Street North, Waterloo ON, N2L 2Y5 Canada}
\furtheraddress
 {Department of Physics, Syracuse University, Syracuse, NY 13244-1130, U.S.A.}
\email{rsorkin@perimeterinstitute.ca}

\AbstractBegins           
Taking as a hypothesis a form of the labour theory of value, 
and {\it\/without assuming equilibrium\/},
we derive an equation that yields the profit-rate $\pi$ as a function of time.  
For a mature economy, $\pi(t)$ reduces to the product of two factors: ($i$) a
certain {\it\/retarded average\/} of the sum of the growth-rates 
of productivity and of the size of the labour-force measured by hours worked,
and ($ii$) the ratio of the current rate of surplus value 
%% (always $\le1$)
to its own retarded average.
We also suggest an empirical test of the equation. 
%
%% a certain retarded average of the rate of surplus value (which cannot exceed unity) with a
%% certain retarded average of the sum of the growth-rates of productivity and of the size of
%% the labour-force measured by hours worked.
%% The derivation holds equally in equilibrium and non-equilibrium conditions. 
%% and is potentially testable empirically. 
\bigskip
\noindent {\it\/Keywords and phrases\/}:  labour-value, profit-rate, time-dependence, non\-equilibrium economics
\AbstractEnds                                

\bigskip

%: Turn on spacing for body of paper

\sesquispace
\vskip -10pt

\section{I.~Introduction}
Any model of capitalist dynamics that rests on an assumption of economic equilibrium can
have at best a limited explanatory value.  Even without having witnessed the upheavals of
the past ten years, I doubt that any serious observer would want to ignore the fact that
{\it\/disequilibrium\/} appears more characteristic of capitalist economies than its
opposite.  The history of capitalism presents us not with smooth development, but with a
series of booms and subsequent ``crises'' interspersed with even more disruptive episodes
like wars (including wars conducted within the capitalist ``core'' and wars inflicted by
the core nations on the ``halo'' regions of the capitalist world).  Faced with this chaotic
panorama, we must ask whether there exist economic models robust enough to survive under
conditions where ``competitive equilibrium'' (if it ever existed) has been lost.

As soon as one decides to seek such a model one comes face to face with a problem.  How are
we to attribute exchange-values to commodities?  Neo-Ricardian approaches assign
values-qua-prices by reference to an input-output matrix or its generalizations.  Other
approaches, even farther removed from reality, appeal to the hypothetical ``utilities'' and
``preferences'' of atomized economic actors.  In both cases, an appeal to uniform growth or
to some other type of equilibrium is essential.  One might attempt to rescue such models by
adding in small fluctuations away from equilibrium, but that could be done only if the
equilibrium could be claimed to be stable, or if it could be claimed to become stable via
some sort of coarse-graining.  As far as I know, such claims have not been put forward with
any seriousness, and no such generalization has been attempted.

In the face of such difficulties, it is tempting to pursue the long-standing idea that
exchange-value derives from labour-value.  According to this hypothesis, known usually as
the ``labour theory of value'', prices are merely a phenomenal form taken by values which
are more fundamentally defined by asking how much accumulated human labour went into
producing a given item.  One can of course raise many questions about how well defined
this labour-value really is, how it relates to prices, etc.  This is not a discussion I
want to enter into here.  Instead I simply want to demonstrate that if one uses
labour-time (denominated in hours, for example) as one's measure of value, and if one defines
profit in terms of labour-values, then one can derive a 
general equation for profit-rate that expresses it as the ``historical resultant'' of
variables that are relatively close to being directly observable, namely the rates of growth
of productivity and of the labour-force, together with the rate of surplus value (or
equivalently exploitation)\footnote{$^\dagger$}
{In this way, the formula offers a natural generalization of the equilibrium result that
 profit-rate is in essence just a measure of the rate of growth of the economy.  See for
 example the Sraffa input-output model in [1] or the von Neumann ``balanced growth''
 ray in [2].}
%.  

If one accepts the labour-value hypothesis, then one would expect that profit defined
thereby would offer a conceptual tool useful for the analysis of capitalist economies.  One
might also go further and expect that ``labour-value profit'' would equate roughly to
profit as measured by prices.  If so then  the underlying assumptions would be opened up
to an empirical test.

\section{II.~Assumptions}            
The analysis to be presented herein will be self-contained, but
%
% However 
it is worth pointing out its relation to the discussion found in volume III of
Das Kapital, in the section on the ``tendentially falling rate of profit''  [3] (which seems to
be Engels' working out of a preliminary draft by Marx).

The argument made there works entirely with labour-values and could perhaps be summarized as
follows, if you will allow me to emphasize those aspects that relate best to the present
analysis.
Profit, insofar as it is invested, implies an increase in the mass of
%% fixed 
constant capital, which in turn raises the ratio of total capital to surplus value.  But, since only
surplus value ends up as profit, the rate of profit must fall secularly.  In a word, present
investment causes future profit-rates to fall.
However, this fall is only a tendency which has to contend with a number of possible
``counteracting causes'' [Entgegenwirkende Ursachen], among them the lengthening of the
working day, an increase in the rate of exploitation, the cheapening of capital due to
rising productivity, and the effects of foreign trade and imperialist super-profits.
%
% state the argument better: ratio of constant/variable rises while surplus value is limited for fixed number of workers.

Apparently the falling rate of profit was not a {\it\/prediction} of the analysis but an
``observed fact'' that writers on economics were attempting to explain.  And indeed, we will
see that the profit rate will fall for some time if it begins at a high value.  
However, we will also see that if we more quantitatively take into account the first three
of the above ``counteracting causes'', as well as the related possibility of growth in the
number of employed workers,
we will arrive at a floor below which the profit-rate need not fall.  On the other hand, we
will also arrive at an upper bound above which --- on average and in the long run --- it
cannot rise.

Our analysis will treat the economy as a whole, that is we will treat everything 
{\it\/in aggregate\/}.  Our main results will be equations (5) and (16).  In
deriving them, we will employ for simplicity a {\it\/continuous-time\/} model of the
economy, i.e. there will be {\it\/flows\/} of raw materials, of labour, etc.  We will also
work in the limit of zero ``latency time'', whence the only capital that will enter into the
profit-rate will be fixed capital.  (With nonzero latency time, other forms of capital like
inventory and ``advanced'' wages would also influence the profit-rate, but some of them
could still be treated as special cases of fixed capital.)
%
%% (This leads to important differences with vol III and many other analyses)
%
Thus we will make (at least) the following assumptions and idealizations.

\item{(1)} A {\it\/closed\/} capitalist economy.  No (or only equal) trade, no destruction of capital (as by war). 

\item{(2)} The rate of profit is equal to total net profit divided by total fixed capital.

\item{(3)}  Profit and capital will be measured by their labour-values, as will all other items.
%% measured and aggregated 

\item{(4)} All profits are invested and there is no saving by workers.

\item{(5)} All investment goes into capital formation.

\section{III.~Some consequences}
Let all values be measured in current labour time (replacement values at time $t$) and let
$K=K(t)$ be the (value of the) total fixed capital at time $t$.  (Recall that we are
treating the economy in aggregate.)
If $\pi = \pi(t)$ is the
{\it\/rate\/} of profit then in unit time the {\it\/amount\/} of profit will be by
definition $\pi K$, whose reinvestment causes the value of fixed capital to {\it\/increase\/} at a
rate $dK/dt=\pi K$.  At the same time however, the cheapening of capital goods {\it\/reduces\/}
$K$ at the rate $dK/dt=-rK$, where $r$ is the rate of productivity growth 
in the capital goods sector.
%
%% Ignore following for now (see 1_comments)
%% \footnote{$^\flat$}
%% %
%% {Under some conditions $-rK$ might actually show up as an accounting loss [R::orig3].
%%  Insofar as this happens $r$ will not be ``counteracting'' and $\beta$ below would
%%  coincide with $\gamma$. }
%
Combining these competing effects yields
$$
      {dK \over dt} = -r K + \pi K = (\pi - r) K              \eqno(1)
$$
If only a fraction $\phi=\phi(t)$ of profit ends up being invested then (1) becomes 
$$
     {dK \over dt} = -rK + \phi \pi K   \ ,        \eqno(1a)
$$
but for simplicity we have assumed that $\phi\ideq1$ (assumption 4 above).  
We have also ignored any destruction of capital, as by war or by scrapping still functional
machinery for lack of demand (assumption 1 above).

To proceed further, define 
$P$ as the number of workers, 
$\lambda$ as the fraction of the day each spends working on  average, 
and $w$ as the wage rate (hours of wage paid to a worker per hour of his/her work).
Recall here that wages are being measured directly
in units of labour-time.  
Accordingly, $w$ and $\lambda$ are both dimensionless numbers between $0$ and $1$.\footnote{$^\star$}
{We neglect those rare circumstances in which $w$ might exceed unity or fall below zero. We
 similarly neglect the possibility (if it is one) that ``speedup'' and overtime become so
 extreme that $\lambda$ effectively exceeds unity.}
The total wages paid per unit time are thus $w\lambda P$, which subtracted from the value
produced, namely $\lambda P$, leaves for the collective capitalist a surplus value $s$ of
$$
     s = \lambda P - w \lambda P                \eqno(2)
$$
It is convenient to combine $\lambda$ and $P$ into the product 
$L=\lambda P$ 
(which is the aggregate labour-time expended per unit time)
and to define
$\sigma = 1 - w$, 
which will be called the {\it\/rate of surplus value\/}, 
and which is also a pure number between $0$ and $1$ representing
the fraction of the workers' production retained by the capitalists.\footnote{$^\dagger$}
{The commonly defined ``rate of exploitation'' $e$ is related to $\sigma$ by $e=\sigma/(1-\sigma)$.}
Combining these definitions yields for the mass of surplus per unit time,
$$
            s = (1-w) \lambda P = \sigma L
$$
Finally, since profit equals $s$ by definition, 
we find for the {\it\/rate\/} of profit, $\pi=s/K$, simply
$$
   \pi\  =\  \hbox{rate of profit} \ =\  {\sigma L \over K}  \eqno(3)
$$
Now how does this labour-value profit-rate change with time?

To answer this question, 
we need only differentiate (3) and combine the result with
(1) to obtain
$$
   {\dot\pi\over\pi} \ =\  {\dot\sigma\over\sigma} + {\dot L\over L} - {\dot K\over K} 
                     \ =\  {\dot\sigma\over\sigma} + {\dot L\over L} + (r - \pi) \ ,
$$
an equation which will look slightly simpler if we define
$$
     \beta\  =\  {\dot L\over L} + r                    \eqno(4)
%%     \beta\  =\  {\dot L\over L} + r  \ =\ {\dot \lambda\over \lambda} + {\dot P\over P} + r
$$
so that it becomes
$$
           {\dot\pi\over\pi} = {\dot\sigma\over\sigma} + \beta - \pi                  \eqno(5)
$$ 
Notice here that 
${\dot L\over L}={\dot \lambda\over \lambda}+{\dot P\over P}$ is just the rate of
growth of the number of workers plus the rate of lengthening of the working day.
Hence (as is anyway obvious) $L=\lambda P$ can rise over long periods only if $P$ does,
since $\lambda$ is bounded above by unity.

One sees how almost all of the ingredients of the ``falling profit-rate'' discussion are
represented in (5).  
If one were to keep only the third term in (5), it would reduce to
$\dot\pi=-\pi^2$, and one would 
deduce that $\pi$ must
decrease forever, falling asymptotically to zero
(or diverging to $-\infty$ if it were negative).
But this would be to neglect the effect
of the two ``counteracting'' terms.  The first of these, namely $\dot\sigma/\sigma$,
can check the fall to the extent that the rate of exploitation increases, but its effect
necessarily dies out as $\sigma$ approaches its maximum value of unity.  The other
counteracting term, namely
$\beta={\dot L\over L}+r={\dot\lambda\over\lambda}+{\dot P\over P}+r$, 
receives three distinct contributions.  The contribution $\dot\lambda/\lambda$ corresponds
to lengthening the working day, but its effect is limited in the same manner as is that of
$\dot\sigma$.
Therefore, if $\beta$ is to be effective for more than a limited time, it can only be
because either $P$ (the size of the workforce) or the productivity of labour (corresponding
to $r$) grows indefinitely.  The net effect is that the long-term profit rate is governed by
these two underlying growth rates.
(Arguably they cannot act forever either, because no exponential growth can last forever.
But the time-scales on which these limitations would assert themselves are evidently much
longer than those belonging to $\sigma$ and $\lambda$.)

In the next section, we will draw some more quantitative consequences from (5).  In
fact (5) will let us compute the profit-rate at any time $t$, provided that we are given the
values of $\sigma$ and $\beta$ at all earlier times.\footnote{$^\flat$}
{If the economy is insufficiently mature, then one needs also the initial value of
 $\pi$ or its equivalent.}
For now, we just note that, if we ignore the transient effects of a changing $\sigma$, the
profit-rate $\pi$ necessarily moves toward $\beta$, decreasing when it exceeds $\beta$ and
increasing if it falls below $\beta$.  In this sense $\beta$ is the ``reference value'' that
$\pi$ always seeks.

Before turning to the general analysis just referred to, let us mention a simple example of
what can be expected. Suppose that both $\sigma$ and $\beta$ are independent of time and that
at $t=0$ there is some labour but no capital in existence (``new economy'').  Integration of
(5) (conveniently done by changing to the variable $z=1/\pi$) yields in this
situation
$$ 
      \pi(t) = {\beta \over 1 - e^{-\beta t}}  \ .    \eqno(6)
$$ 
Thus the profit-rate under these conditions is infinite at first, but it falls to its
equilibrium value of $\beta$ on a time-scale whose duration $\beta^{-1}$ is itself set by
$\beta$, i.e. by the underlying rate of growth of productive capacity.
The ``tendency to fall'' and the ``counteracting causes'' are then in balance.

\section{IV.~ The profit rate more generally}           
In general neither $\beta$ nor $\sigma$ will be constant.  Nevertheless, it is clear from
the fact that (5) is a first-order differential equation that given the initial
value of $\pi$, we can, within the limits of our simplified model, deduce $\pi(t)$ for all
$t$ if we know the history of $\beta$ together with that of $\sigma$.
By working out this dependence explicitly, we will arrive at the general solution for
$\pi(t)$ recorded in equation (16).  In consequence of this equation, we will also
see that the initial conditions tend to drop out in a ``mature'' economy, leading to the
result that the profit-rate at a given time is the product of two averages which summarize
the historically recent values of $\beta$ and $\sigma$, respectively.  We will also derive
an upper bound on $\pi$ that is entirely independent of initial conditions, and is
relatively independent of $\sigma$ as well.
%% given then by the average $\beta$ over relatively ``recent'' times.

Our task, then, is to solve (5), presupposing that both $\beta$ and $\sigma$ are
known functions of time.  As it stands, (5) contains a term quadratic in $\pi$, but
we can render it linear by working with $1/\pi$ rather than $\pi$ itself.  In fact 
the new variable $Z$ defined by 
$$
         Z   \ =\  \sigma / \pi            \eqno(7)
$$
will be a slightly more helpful choice.  From equation (3) we see that 
$Z=K/L$ measures the fixed capital per worker hour.  It is thus closely related to what is
sometimes called the ``organic composition of capital''.
Rewritten in terms of $Z$ equation (5) becomes simply
$$
         {dZ \over dt} + \beta \; Z \ =\ \sigma           \eqno(8)
$$
Thus $\beta$ acts like a ``restoring force'' that (when positive) pulls $Z$ toward $0$,
while $\sigma$ acts like a ``source'' that pushes $Z$ toward higher values. (More
exploitation implies more rapid accumulation of surplus value.)  At any rate, we have in
(8) a linear equation 
whose general solution is
%% which can be solved by quadratures to yield
%
$$
     Z(t) = Z_0 \ e^{-B(t,0)} \ +\  \int\limits_0^t e^{-B(t,s)} \ \sigma(s)\  ds     \eqno(9) 
$$
where
$$
    B(t_2,t_1) = \int\limits_{t_1}^{t_2} \beta(t)\ dt      \eqno(10)
$$
and $Z_0=Z(t=0)$ is the initial value of $Z$.  
(One can verify this solution by direct substitution into (8).)
From $Z(t)$ we can recover $\pi(t)$ 
trivially,
%% by simple algebra, 
but first let us
interpret the expression $B(t_2,t_1)$ and the integrals in which it occurs.

To that end, notice first that since $e^{-B}$ is always positive the integral in
(9) constitutes a weighted sum of $\sigma$ with recent values counting most
heavily (unless the economy is shrinking!) and very early values being exponentially damped.
Appropriately normalized
this integral
therefore defines a certain kind of ``moving average'' of $\sigma$ 
(or of any given function of time) 
which I'll denote by $\ave{ \cdot }$~:  
$$
   \ave{ \sigma(t) } \ =\ \int\limits_0^t e^{-B(t,s)}\  \sigma(s) \ ds  \ \bigg/ \ \int\limits_0^t e^{-B(t,s)}\ ds  
    \eqno(11)
$$ 
The denominator of (11) can also (though less obviously) be interpreted as an
average, this time of $\beta$ itself.  
To that end let us define $\betabar=\betabar(t)$ through the equation
$$
    { 1 - e^{-t\betabar(t)} \over \betabar(t) } = \int\limits_0^t e^{-B(t,s)}\  ds  \ .   \eqno(12) 
$$
Then $\betabar$ is a kind of ``self-weighted moving average'' of $\beta$ in the following sense (see Appendix):
\item{(1)} For each $t$, equation  (12) defines $\betabar(t)$ uniquely.
\item{(2)} If $\beta$ is the constant function $\beta_0$ then $\betabar=\beta_0$ \ .
\item{(3)} If (for all $t$) $\beta_1\le\beta_2$ then $\betabar_1\le\betabar_2$\ ; 
           in particular any (lower or upper) bound for $\beta$ is also one for $\betabar$~.

\noindent
Moreover this second average is closely related to the first one, 
since, as proven in the appendix,
$$
   {\betabar \over  1 - e^{-t\betabar}} \ =\ { \ave{\beta} \over 1 - e^{-B(t,0)}}   \eqno(13)
$$

In order to bring out further the intuitive meaning of the weighting factor $e^{-B}$ which
defines the average $\ave\cdot$,
let us observe that according to (10) and (4), $B(t_2,t_1)$ is the sum of the
integrated growth-rates of $L$ and of productivity.  If we temporarily introduce a symbol $\Pi$ to
represent productivity and define $Q=L\,\Pi$ then $\beta=\dot{L}/L+\dot\Pi/\Pi$ is the rate of
change of $\log{Q}$ and its integral $B$ is just the logarithm of $Q(t_2)/Q(t_1)$.  
Thus 
$$
    e^{B(t,s)} = Q(t) / Q(s)                   \eqno(14)
$$ 
is a
kind of measure of how much the productive capacity of the economy has grown between $t_1$
and $t_2$, and the effect of our weighting factor $e^{-B(t,s)}$ 
in the above integrals
is to discount the contribution from a
given historical moment in proportion as the productive capacity then was smaller.  In a
growing economy, such an average damps out the contribution from early times.

Thanks to these observations, 
we can also express the average (11) as
$$
   \ave{\sigma(t)} \ =\ \int\limits_0^t Q(s)\ \sigma(s) \ ds  \ \bigg/ \ \int\limits_0^t Q(s)\ ds  
    \eqno(11a)
$$ 
while the corresponding recasting of $\betabar$ appears as
$$
    { 1 - e^{-t\,\betabar(t)} \over \betabar(t)}\ =\ \int\limits_0^t Q(s)\  ds \  \big/ \ Q(t)
    \ .   \eqno(12a) 
$$

With these definitions in mind, we can assemble equations (9), (14),
(11a), and (12a) to obtain [where $Z=Z(t)$, $Q=Q(t)$, $\betabar=\betabar(t)$]
$$
        Z = {Z_0 Q_0 \over Q}+ \ave\sigma {1-e^{-\betabar t}\over \betabar}
$$
which with the aid of the identity (13) becomes
$$
        Z = {Z_0 Q_0 \over Q} + \ave\sigma {1 - Q_0/Q \over \ave\beta}
$$
or after rearrangement,
$$
            Z \ =\ { \ave\sigma \over \ave\beta} 
              \ +\  {Q_0\over Q} \left[Z_0 - { \ave\sigma \over \ave\beta} \right]  \eqno(15)
$$ 
This is our main result expressed in terms of $Z=\sigma/\pi$.  In a sufficiently mature
economy (one which has already expanded by a large factor) $Q_0/Q$ will be small compared to
unity and we will be left with the approximate equality, $Z=\ave\sigma/\ave\beta$.

Finally, we can substitute the definition (7) of $Z$ to obtain an equation for the
profit-rate at time $t$ which, after a small rearrangement, 
furnishes our main equation for $\pi$: 
$$
    \pi \ =\ 
     {\ave\beta \ \sigma/\ave\sigma 
      \over 
      1 + {Q_0\over Q}\left[ {\ave\beta\over\pi_0}{\sigma_0\over\ave\sigma} -1 \right]}
     \eqno(16) 
$$
Once again the term containing $Q_0/Q$ can be dropped in a mature economy and we are left
in that case with the approximate equality
$$
   \pi \ =\  \ave\beta \ {\sigma\over\ave\sigma}    \eqno(17)
$$ 
Even without making any assumption about maturity, we deduce from (16) the
inequality
$$
    \pi \ \le \ \ave\beta \ {\sigma\over\ave\sigma} \ (1-Q_0/Q)^{-1}   \eqno(18)
$$
which bounds the profit-rate above by an expression not much bigger than (17), and
bounds it even more precisely by that value times $(1+Q_0/Q)$, a factor close to unity which
should provide an adequate approximation in any but a very young economy.

What the last three equations tell is that, 
apart from initial transients and temporary fluctuations, 
the profit-rate coincides with 
the growth-rate $\beta$ averaged over historically recent times.  
They also tell us that the fluctuations about this average 
come from the factor
${\sigma/\ave\sigma}$
%% ${\sigma\over\ave\sigma}$, 
which will be appreciable only when 
$\sigma$ 
(the rate of surplus value)
rises sharply above, or falls sharply below, its historically recent average $\ave\sigma$.
And since $\sigma$ is in any case less than 1, 
a sharp increase is possible only if it was recently much less than unity.

%% And this leads immediately to the inequality
%% $$
%%      Z(t)  \ge  \int\limits_0^t e^{-B(t,s)} \sigma(s) \, ds \ .   \eqno(E::orig14)
%% $$

%% With the definitions (E::orig15) and (E::orig16), (E::orig14) becomes
%% %
%% $$
%%    Z \ \ge \  {1 - e^{-t\betabar}  \over  \betabar}  \ \ave{f} \ , 
%% $$
%% %
%% which implies, with the help of (E::orig3 a) and (E::orig8 b),
%% $$
%%    \pi \ \le \ {\betabar  \over  1 - e^{-t\betabar}} \   {\sigma \over  \ave{\sigma} } \ . \eqno(E::orig18)
%% $$
%% %
%% This is the first form of our desired general upper bound.  A somewhat simpler version
%% follows (for $\betabar\ge0$) from the general inequality (see the Appendix),
%% $x/(1-e^{-x})\le 1+x$~:
%% %
%% $$
%%           \pi \ \le \  (\betabar + t^{-1}) \; \sigma/\ave{\sigma} \ .    \eqno(E::orig19 a) 
%% $$
%% %
%% A final version results from using (E::orig17) together with the approximation  $1-e^{-B(t,0)}\approx1$~:
%% $$
%%      \pi \  \le \  \ave{\beta} \  \sigma/\ave{\sigma} \ .    \eqno(E::orig19 b)
%% $$

\section{V.~Comments and Extensions}
Under conditions of equilibrium, one would expect the profit-rate to reflect directly the
underlying growth-rate, because the latter characterizes the entire process, and one would
expect to recognize this same rate no matter what variable one chose to monitor.  
It is therefore
no surprise that equilibrium models 
(``neo-Ricardian'' and others) 
produce the result $\pi=\beta$.  In
disequilibrium conditions, however, some variables will be increasing while others decrease,
fluctuations will be significant, if not dominant, and the reasoning behind the equilibrium
results will fail.  It is thus interesting that, at least in the simple model of this
paper, the profit-rate is still governed to a large extent by the growth rates, $r$ and
$\dot{L}/L$ ($\beta$ being their sum).  The difference however is that the contemporaneous
value of $\pi$ is no longer tied to the contemporaneous value of $\beta$, but to its
historical average $\ave\beta$
taken in a precisely defined sense.  Moreover, significant short-term
fluctuations in $\pi$ will in general take place, depending on the contemporaneous rate of
surplus value $\sigma$ in relation to {\it\/its\/} historical average $\ave\sigma$.   In
a relatively young economy the initial conditions will be important as well.
(One can of course set the initial time whenever one wants.  The equations will
still hold.)

\subsection{An illustration}
By way of illustrating our conclusions, let us imagine a closed economy whose initial time,
$t=0$, is chosen to be around the end of the last ``world war'', so that at present
$t=70yr$, and suppose that its ``productive capacity'' $Q$ has expanded since then by a factor
of ten: $Q/Q_0=10$.  The historical averaging that enters into our equations would then
weight recent values of $\beta$ and $\sigma$ about 10 times more heavily than values from 70
years ago.  The latter could thus still make some difference, but not too much.
Let us suppose further that $\beta$ recently has hovered around 2\% per year, having been
bigger previously such that currently $\ave\beta=0.025/yr$, 
and that the average rate of exploitation has reached $\ave\sigma=0.50$.  
Let us also suppose that in the past couple of
years the capitalists, in a desperate attempt to raise profits, have driven $\sigma$ to its
maximum possible value of unity: $\sigma=1$ (which of course they could never really attain).
The inequality  (18) would then allow at present a current labour-value profit-rate
of at most
$$
        0.025 / yr \times {1.0\over 0.5} \times {1 \over 1 - 0.1} = 0.056 /yr
$$
In other words, the annual profit rate could not currently exceed 5.6\% even if the workers
were made to ``live on air''.  

Even this excess over $\ave\beta$ would only be temporary. 
In fact one can estimate in general 
that $\pi(t)$ can not exceed $\beta(t)$ by an amount $\eps$ for a time
much longer than $1/\eps$, or somewhat more precisely, this time multiplied by
$\ln(1/\sigma)+\ln(1/(\beta'+\eps))$~,
~$\beta'$ being the value of $\beta$ at the end of the period.

%%  increase With growth rates of one per cent per year or higher and times $t=100$ years or more,
%% $t^{-1}$ can be disregarded in (E::orig19 a); and (E::orig19 b) should also hold.  So for
%% present-day advanced capitalist economies (insofar as they are closed systems!) only an
%% increase in $f=e\lambda$ above its average $\ave{f}$ could raise $\pi$ above $\ave{\beta}$.
%% Since $e,\;\lambda\le1$, such an increase can only be temporary, and the greater
%% $e\lambda$ is to begin with, the less the amount by, and the shorter the time over which,
%% $f$ can exceed $\ave{f}$.  [A time of about $\beta^{-1}\lg f^{-1}$ seems to be the upper
%% limit.]  In other words, an increase in the workday or in the rate of surplus value can be
%% ``counteracting'' only for a time not exceeding $\beta^{-1}$ and only if the original
%% workday is very short or the rate of exploitation very low.  (Of course $\beta^{-1}$ can
%% be quite long if the growth rate is small.)

\subsection{Further remarks on equilibrium models}
%
%% One assumption we have {\it\/not\/} made is that of equilibrium. Why, in this paper, have we been at pains
%
I have already expressed more than once herein, 
the view that equilibrium models are not to
be trusted.  Perhaps the most devastating evidence for this is the observation that
equilibrium is simply not the historical rule, but one can also give more ``technical''
reasons why a hypothetical equilibrium profit rate is likely to be wrong.  For this rate is
essentially the maximum possible rate, a statement that is
 in some sense the content of the Frobenius-Perron theorem, which states
 in particular that for any set of prices there will always be at least one
 industry in which profitability is at or below its equilibrium value.  In itself
 this is a fairly weak limitation, but presumably one could show that any growth
 trajectory far from the equilibrium one would soon encounter severe shortages and
 hence severe ``realization'' problems, resulting in markedly lower profits.  
 A fuller analysis of this situation might also be able to elucidate the loss of profit
 inherent in any {\it\/return\/} to equilibrium after new techniques or products
 have been introduced.
In any case, if the equilibrium rate is an upper bound, it follows that slumps can lower
profits below this hypothetical value more than booms can raise them.  Thus, even if prices
and product-mixes averaged out to their equilibrium values, the (nonlinear) relation between
them and the (overall) profit rate would in general cause the latter to deviate
systematically from its own equilibrium value.  Similarly, any nonlinearity in the response
of investors to fluctuating prices and markets would invalidate arguments of the
Okishio-Theorem type, that employ a notion of capitalist rationality defined solely with
respect to the current {\it\/equilibrium\/} price-vector
(as in [4] for example).

\subsection{Possible extensions}
In deriving our main results, equations (5) and (16), we have relied on several
assumptions and idealizations.  Some of them could easily be relaxed.  For example one could
accommodate a situation where not all surplus value resulted in capital formation by using equation
(1a) in place of (1),
with $\phi$ chosen to take account of dividends and personal consumption by capitalists.
%
%% In a simple model, we might then posit for $\phi$ an
%% equation like
%% $$
%%    \phi = (1 - \delta) (1 - u) + u / \sigma \ ,  \eqno(E::orig20)  
%% $$
%% where $\delta$ = dividends/profits and $u$ = (personal) savings/ (personal) income.

The assumption that capital is not destroyed could also be relaxed by incorporating into the
variable $r$ an additive contribution representing the rate of its destruction.  Such an $r$
could no longer be interpreted simply as a growth-rate, but mathematically the above
analysis would go through as before.  Equations like (11a) and (12a)
would become incorrect, but re-expressed in terms of $B(t,s)$ via (14), all our
final results would remain the same.

Some of our other idealizations would be more difficult to do without, notably the
assumption that the economy in question is a closed system.  One might also wonder whether 
(3) was a good approximation in a more service-based economy.  Similarly,
one might question whether, or how, ``fictitious'' financial capital ought to be included in
the variable $K$, or how to take into account labour which in some sense is ``unproductive''.
%% [[add some examples?: security guards, internet trolls]] 

\subsection{Possible tests}
Beyond these more ``technical'' questions are the conceptual questions raised by the use of
labour-values.  In this paper, we are simply adopting labour-values as our starting point
and drawing out the consequences mathematically.  The advantages of doing so are first of
all that we avoid the equilibrium fiction (if I can call it that), and secondly that one can
give at least a fairly clear definition of the crucial parameter $r$ measuring productivity
growth.  One also has the feeling that labour values tap into something basic about the
economy which empirical prices do not.\footnote{$^\star$}
{The observation seems relevant here that labour-power is the one commodity which (ignoring
 slave markets and analogous exceptions) is not produced for sale by capitalists.}
For this reason $\pi$ might also have a certain
analytical interest in its own right.
(A good question is why economists worry so much about ``productivity''
if labour values are irrelevant.)

%% The remaining way to raise profits would be to raise $\beta$ itself, either by
%% technological or organizational improvements fostering greater productivity growth $r$, or
%% by increases in $\gamma$ driven for example by immigration or the entry of a larger
%% fraction of the population into the workforce (as has happened e.g. with more women
%% working for pay).  In the longer run however only basic research or actual population
%% increase seem capable of sustaining $\beta$.

In this connection, one might wonder whether some other ``universal'' measure of value could
have been used in place of labour-time.
Indeed, the fact that $\beta$ represents just the growth rate of productive capacity (more
precisely of the hypothetical capacity for turning out capital goods if all available labour
were devoted to them) suggests that some form of the equations relating $\pi$ with $\beta$
might be provable without recourse to any theory of value at all.  However any attempt to
define $\beta$ in this way (not to mention $\sigma$) encounters the ambiguity inherent in
defining growth rates in the face of the continual emergence of new commodities and
techniques and the disappearance of old ones which economic growth entails.
The present treatment, based on assigning to each commodity its labour-value, renders the
crucial variable $\beta$ unambiguous to a significant extent.  Nevertheless, it seems clear
that of the two factors entering into its definition, namely hours worked and productivity,
the latter is less well defined than the former.

From an empirical standpoint the question would be whether, and to what extent, the
phenomenal profit rate can be identified with the labour-value profit-rate studied in this
paper.  Some economists seem to think that it can, many others think not.  At this stage of
economic theory, 
I believe one must treat
either answer
%% it 
simply as 
a hypothesis that one can adopt (or not) as the basis of further analysis, 
just as one does with hypotheses in the physical sciences.
Personally, I have very little idea how to address this issue theoretically, but our results
above provide the beginnings of a way to address it empirically.  (Beginnings because some
of the simplifications we have made would need to be either corrected for or corroborated.)
To the extent that $r$, $\pi$ and $\sigma$ can be observed ($L$ unquestionably can), one
could compute (from (16) or (5) together with $\beta$ and $\sigma$) $\pi$ as
a function of time and compare the resulting graph with the data.
Or if this test were too difficult because $\sigma(t)$ was too hard to observe reliably
given available statistics, one could as a second-best test, deduce $\sigma(t)$ vice versa
from (5) together with $\pi$ and $\beta$, and see whether the resulting graph was at
least intuitively plausible.

%% and then defining $r(t)$ as the rate of decrease of the {\it\/value\/} of the
%% capital stock existing at time $t$.  

%% On the other hand any other universally applicable concept of value would have served this
%% purpose equally well.  Indeed the ``true'' measure of profit may ultimately depend on why
%% one wants to measure it, and in particular on what one thinks the investment decisions of
%% capital actually seed to maximize.

%% Two perennial and interrelated economic controversies attend the labor
%% theory of value and the ``law of the tendency of the rate of profit to fall''.  
%% Although {\it\/Das Kapital\/}'s original discussion of the latter [R::orig1] was
%% based on the former, some of the more recent treatments [R::orig2] proceed instead
%% from a ``neo-Ricardian'' theory of prices.  However it seems risky to
%% base any assertion about secular trends on a theory which presupposes an
%% economy in competitive equilibrium.

%% Add comment that labour-value not produced for sale? done now in footnote above! 

\section{Appendix.  Some technical details}           
%% \section{Appendix. Properties of $\betabar$ and other technical details}           
%
In terms of the function
$$
         F(x) \ \ideq\ {1 - e^{-x} \over x} \ =\ \int\limits_0^1 e^{-xs} ds \ , \eqno(\Amoja) 
$$
the definition (12) of $\betabar$ reads
$$
       F(\betabar) = \int\limits_0^t e^{-B(t,s)} ds\ .       \eqno(\A2) 
$$
We want to show that $\betabar$ is thereby well-defined and also to establish the further
assertions made in the text.

In the first place notice that, because the integrand in (\Amoja) is a monotone decreasing
function of $x$, $F$ is also monotone decreasing.  Moreover it is clear that
$F(-\infty)=+\infty$, $F(+\infty)=0$, so that $F$ is a one-to-one map of the reals onto
the positive reals.  Hence (\A2) has a unique solution $\betabar$; which was assertion (1)
of the text.

Now if $\beta=\beta_0$ is constant then from (10), $B(t,s)=(t-s)\beta_0$, whence
the integral in (\A2) is
$$
    \int\limits_0^t e^{-\beta_0(t-s)} ds = { 1 - e^{-\beta_0t}\over \beta_0} \ .
$$
Then the unique solution of (\A2) is obviously $\betabar=\beta_0$~, which was assertion (2).

Given this, the second part of assertion (3) follows from the first.  To prove the first
simply notice that if $\beta$ increases pointwise then $\int e^{-B(t,s)ds}$ decreases, so that
$\betabar$ in (\A2) must increase because $F$ is monotone decreasing.

Finally let us demonstrate the identity (13).  To that end, let us first
write out $\ave{\beta(t)}$ in the form (which follows immediately from (11) and
(12)),
$$
    \ave{\beta } \ =\ { \betabar   \over 1 - e^{-t\betabar} } \ \int\limits_0^t e^{-B(t,s)}\  \beta(s) \ ds \ 
   \eqno(19)
$$
Next observe that 
$\partial/\partial{s} \ e^{-B(t,s)} = \beta(s) \, e^{-B(t,s)}$,
and substitute this into (19) to obtain 
$$
  \ave{\beta} \ =\ 
  {\betabar \over 1 - e^{-t\betabar}} \ \int\limits_0^t ds \; {\partial\over\partial{s}} \; e^{-B(t,s)}
  \ =\  
  {\betabar \over 1 - e^{-t\betabar}} \ \big(1 - e^{-B(t,0)} \big)  \ ,
$$
from which (13) follows immediately.

%% Finally we must prove the inequality
%% $$
%%           {x \over 1 - e^{-x}} \ \le \  1 + x \qquad (x\ge0)
%% $$
%% that was used to derive eq. (E::orig19 a).  For $x\ge0$ we have
%% $e^x$ = $1+x+x^2/2+\cdots\ge1+x$ $\ \implies\  e^x\le(1+x)^{-1}$
%% $\ \implies\  1-e^{-x}\ge1-(1+x)^{-1}=x/(1+x)$
%% $\ \implies\  x/(1-e^{-x})\le 1+x$ as required.

%% SAMPLE FIGURE

%:  figure 1

%% \vskip 1.1cm       % <---------- Adjust so that the top of the diagram doesn't blot out any text!
%% \vbox{\bigskip
%%   \centerline {\includegraphics[scale=0.5]{blah.eps}}
%%   \Caption{{\it Figure @@.}  Blah blah blah}}

%: Acknowledgements                           

% (also thank fqxi if used `alisp')
% (also thank OLAM [peyresq] if appropriate)

\bigskip
\noindent
This research was supported in part by NSERC through grant RGPIN-418709-2012.
This research was supported in part by Perimeter Institute for
Theoretical Physics. Research at Perimeter Institute is supported
by the Government of Canada through Industry Canada and by the
Province of Ontario through the Ministry of Economic Development
and Innovation.  

\ReferencesBegin                             

% (ref-equivalence:: )                       
% After the double colon put: tag-1 tag-2 tag-3 ...

\ref [1] Piero Sraffa, {\it Production of Commodities by Means of Commodities: Prelude to a Critique of Economic Theory} 
   (Cambridge University Press, 1960)

\ref [2] J. von Neumann, ``{\"U}ber ein {\"o}konomisches Gleichungssystem und ein Verallgemeinerung des Brouwerschen Fixpunktsatzes'', 
\journaldata{Ergebnisse eines Mathematischen Kolloquiums}{8}{73-83}{1937},
translated as ``A Model of General Equilibrium'', \journaldata{Review of Economic Studies}{13}{1-9}{1945-46}

\ref [3] Karl Marx / Friedrich Engels, {\it\/Das Kapital, Band 3\/}, 3. Teil: Gesetz des tendenziellen Falls der Profitrate,
\linebreak
retrieved from http://ciml.250x.com/archive/marx\_engels/german/kapital3.pdf;
translated in  K.~Marx (ed. F.~Engels), {\it\/Capital vol III\/} (N.Y. International Publishers, 1967) Part~III.

%% \ref [R::orig2] 
%%  P. van Parijs, ``The Falling Rate of Profit Theory of Crisis \dots'',
%% \journaldata{RRPE}{12 (no.1)}{1}{1980}, and references therein;
%% \sepref

\ref [4] J.E. Roemer, ``Continuing Controversy on the Falling Rate of Profit: Fixed Capital And Other Issues'', 
\journaldata{Cambridge J. Econ.}{3}{379}{1979}

%% \ref[R::orig3] J. Alberro and J. Persky, ``The Dynamics of Fixed Capital Revaluation and Scrapping'',
%% \journaldata{RRPE}{13 (no.2)}{32}{1981}

\end 
%: Outline mode stuff (put here so doesn't need to be commented out)

(prog1 'now-outlining
  (Outline* 
     "\f"                   1
      "%------"             1
      "%:  "                2
      "%:: "                2       ; now deprecated
      "%: "                 1       
      "\\Abstract"          1
      "\\section"           1
      "\\subsection"        2
      "\\appendix"          1       ; still needed?
      "\\ReferencesBegin"   1
      "% (ref-equivalence"  2
      "\\ref "              2
      "\\end